\documentclass[12pt]{article}

\usepackage{sbc-template}

\usepackage{graphicx,url}
\usepackage{environ}
\usepackage{amssymb}
\usepackage{amsmath}
\usepackage{amsthm}
\usepackage{makecell}
\usepackage{multicol}
\usepackage{hhline}
\usepackage{multirow}
\usepackage{color,colortbl}
\usepackage[ruled,onelanguage,linesnumbered,english,vlined]{algorithm2e}
\usepackage{hyperref}
\usepackage[utf8]{inputenc}

\newcommand\blfootnote[1]{%
  \begingroup
  \renewcommand\thefootnote{}\footnote{#1}%
  \addtocounter{footnote}{-1}%
  \endgroup
}

\definecolor{Gray}{gray}{0.92}

\title{Heuristics based on Adjacency Graph Packing for DCJ Distance Considering Intergenic Regions}

\author{Gabriel Siqueira\inst{1}\and
Alexsandro Oliveira Alexandrino\inst{1}\and\\
Andre Rodrigues Oliveira\inst{2}\and
Zanoni Dias\inst{1}
}
\address{Instituto de Computa\c{c}\~{a}o, Universidade Estadual de Campinas (Unicamp),\\ Campinas, Brazil\\
\nextinstitute
Faculdade de Computação e Informática, Universidade Presbiteriana Mackenzie,\\ São Paulo, Brazil\\
\email{$\{$gabriel.siqueira,alexsandro,andrero,zanoni$\}$@ic.unicamp.br}
}

\makeatletter
\newcommand{\REVIEW}[0]{}

\newcommand{\probleminput}[1]{\gdef\@probleminput{#1}}
\newcommand{\problemtask}[1]{\gdef\@problemtask{#1}}
\NewEnviron{prob}[1]{%
\begin{center}
\fbox{%
\begin{minipage}{0.9\linewidth}
\raggedright
{\centering\scshape #1\par}%
	\parskip=1ex
	\BODY
	\textbf{Input:} \@probleminput \\
	\textbf{Goal:} \@problemtask \\
\end{minipage}
}
\end{center}
}
\makeatother

\begin{document}

\maketitle

\begin{abstract}
In this work, we explore heuristics for the Adjacency Graph Packing problem, which can be  applied to the Double Cut and Join (DCJ) Distance Problem. The DCJ is a rearrangement operation and the distance problem considering it is a well established method for genome comparison. Our heuristics will use the structure called adjacency graph adapted to include information about intergenic regions, multiple copies of genes in the genomes, and multiple circular or linear chromosomes. The only required property from the genomes is that it must be possible to turn one into the other with DCJ operations. We propose one greedy heuristic and one heuristic based on Genetic Algorithms. Our experimental tests in artificial genomes show that the use of heuristics is capable of finding good results that are superior to a simpler random strategy.\blfootnote{This work appeared in the Proceedings of the XVII Brazilian Symposium on Bioinformatics (BSB'2024), \href{https://doi.org/10.5753/bsb.2024.245554}{https://doi.org/10.5753/bsb.2024.245554}.}
\end{abstract}

\section{Introduction}

{\REVIEW
In biology, it is often important to have metrics to compare genomes of different individuals. Such metrics can be used to infer evolutionary distance for the construction of phylogenetic trees. These metrics can also help in the identification of ortholog genes (genes separated by speciation).

Many metrics for genome comparison were proposed over time, including the well established rearrangement distance~\cite{2009-fertin-etal}. The rearrangement distance is a measure of the number of rearrangement operations, large scale mutations affecting the order and orientation of the genetic material, needed to transform one genome into another. One of the most studied rearrangement operations is the Double Cut and Join (DCJ). The DCJ operation consists of cutting a genome in two points and joining the resulting parts.
}

Initially, the DCJ distance was studied in genomes with a single copy of each gene, and the orientation of the genes was taken into account~\cite{2005-yancopoulos-etal}. In that scenario, there is an exact polynomial time algorithm to compute the distance in linear time~\cite{2006-bergeron-etal}. A generalized version of the DCJ Distance Problem considers any two genomes as long as they have the same set of genes. In this case, the DCJ Distance Problem is NP-hard~\cite{2015-shao-etal} and some proposed solutions for it include an integer linear programming formulation~\cite{2015-shao-etal} and an $O(k)$-approximation algorithm~\cite{2017-rubert-etal}, where $k$ is the maximum number of copies of a gene in the genomes.

In more recent works, a new approach was proposed to include information about intergenic regions~\cite{2017-fertin-etal,2020a-brito-etal,2021b-oliveira-etal,2024-oliveira-etal}. The usual representation in these works considers the number of nucleotides between genes, which is called the size of the intergenic region between these genes. With this representation, the DCJ Distance Problem is already NP-hard, even if the genomes have a single copy of each gene and there is a $4/3$-approximation algorithm for it~\cite{2017-fertin-etal}.

The main structure proposed to deal with the DCJ Distance Problem is the adjacency graph. This graph was initially proposed for the problem without gene repetition~\cite{2006-bergeron-etal} and is capable of representing multiple chromosomes, that can be linear or circular. This structure was later adapted to deal with multiple gene copies~\cite{2015-shao-etal,2021c-siqueira-etal} or intergenic regions~\cite{2017-fertin-etal}. However, as far as we know, there is still no work combining multiple genes and intergenic regions for the DCJ Distance Problem.

This work proposes heuristics based on the adjacency graph for the DCJ Distance Problem considering genomes with repeated genes and taking into account intergenic regions and gene orientation. The heuristics are capable of dealing with genomes containing multiple circular or linear chromosomes. We assume that the genomes have the same set of genes. The next section introduces some definitions and formally states the problems. In Section~\ref{sec:heur}, we describe the developed heuristics. In Section~\ref{sec:exp}, we describe some experimental tests, and we conclude the paper in Section~\ref{sec:conc}.

\section{Definitions}

\begin{figure}
    \centering
    \includegraphics[width=0.9\linewidth]{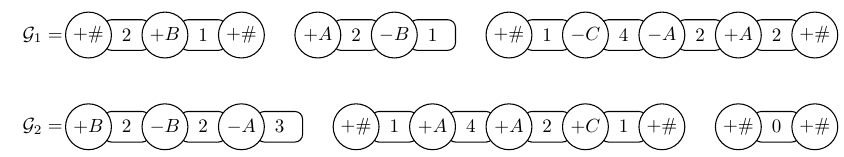}
    \caption{Two balanced genomes $\mathcal{G}_1$ and $\mathcal{G}_2$. The chromosomes of $\mathcal{G}_1$ are represented by the following pairs of strings and list of integers $([+\#,+B,+\#],[2,1])$, $([+A,-B],[2,1])$, and $([+\#,-C,-A,+A,+\#],[1,4,2,2])$. The chromosomes of $\mathcal{G}_2$ are represented by the following pairs of strings and list of integers $([+B,-B,-A],[2,2,3])$, $([+\#,+A,+A,+C,+\#],[1,4,2,1])$, and $([+\#,+\#],[0])$. The linear chromosomes are capped and one chromosome was included in $\mathcal{G}_2$ to ensure the same number of chromosomes in both genomes.}
    \label{fig:genomes}
\end{figure}

We represent a genome as a set of chromosomes. Each \emph{chromosome} $\mathcal{C}$ is encoded by a pair $(S,\breve{S})$, composed of a string $S$ of size $|S|$, representing the genes, and a list of non-negative integers $\breve{S}$ of size $|\breve{S}|$, representing the intergenic regions. Each character of $S$ has an associated sign $+$ or $-$ to represent the orientation of the correspondent gene. We use the term \emph{label} to the symbol used to represent a character, disregarding the sign. Genes that are considered equal will be represented by characters with the same label. Our representation will allow for both linear and circular chromosomes. Figure~\ref{fig:genomes} shows two genomes.

In linear chromosomes, we apply a process called \emph{capping} for the representation. In this process, we insert a character $+\#$ at the beginning and at the end of $S$. We call these characters \emph{caps}. They do not correspond to real genes, but to simplify our definitions they are treated as genes at the beginning and end of the chromosome. Considering this process, if $\mathcal{C}$ is linear, than $|S| = |\breve{S}| + 1$. If $\mathcal{C}$ is circular, then $|S| = |\breve{S}|$.

The $i$-th character of $S$, denoted by $S_i$, represents the $i$-th gene of the chromosome, if it is linear. If the chromosome is circular, we list the genes by cutting it at some point and assume that $S_1$ and $S_{|S|}$ are adjacent. The $i$-th integer of $\breve{S}$, denoted by $\breve{S}_i$, represents the size of the intergenic region between $S_{i}$ and $S_{i+1}$. In circular chromosomes, the integer $\breve{S}_{|\breve{S}|}$ represents the size of the intergenic region between $S_{|S|}$ and $S_1$.

Two genomes are called \emph{balanced} if all labels, except $\#$, appear in the same number of genes, and the sum of the intergenic region sizes is the same in both genomes. In this work we only consider balanced genomes. Additionally, until the genomes have the same number of chromosomes, we add in the genome with fewer chromosomes linear chromosomes with two caps and an intergenic region of size $0$ between them. With this addition of chromosomes the genomes remain balanced and have the same number of characters with the label $\#$.

Given two genomes $\mathcal{G}_1$ and $\mathcal{G}_2$, the \emph{adjacency graph} $\mathbb{AG} (\mathcal{G}_1,\mathcal{G}_2)$ is composed of the vertex set $V$, edge set $E$ (separated in two subsets $E_r$ and $E_d$), and weight function $w:E_r \to \mathbb{N}$. For each character $S_i$ in a string $S$ from a chromosome of $\mathcal{G}_1$ or $\mathcal{G}_2$, we have two vertices $v_{S_i}^t$ and $v_{S_i}^h$ in $V$, if $S_i$ is not a cap, or we have one vertex $v_{S_i}$ in $V$, if $S_i$ is a cap. To simplify our examples, we will use the labels to represent the vertices (adding $h$ and $t$ accordingly). The edges in $E_r$ are called \emph{reality edges} and connect vertices of two consecutive characters $S_i$ and $S_{i+1}$ as follows:
\begin{itemize}
	\item $v_{S_i}^h$ is connected with $v_{S_{i+1}}^t$, if both have sign $+$.
	\item $v_{S_i}^t$ is connected with $v_{S_{i+1}}^h$, if both have sign $-$.
	\item $v_{S_i}^h$ is connected with $v_{S_{i+1}}^h$, if $S_i$ has sign $+$ and $S_{i+1}$ has sign $-$.
	\item $v_{S_i}^t$ is connected with $v_{S_{i+1}}^t$, if $S_i$ has sign $-$ and $S_{i+1}$ has sign $+$.
	\item If $S_i$ or $S_{i+1}$ is a cap, just consider the above cases without the $h$ or $t$ in the vertex correspondent to this cap.
\end{itemize}

The edges in $E_d$ are called \emph{desire edges} and connect the vertices from a character $S_i$ of a string $S$ in a chromosome of $\mathcal{G}_1$ to the vertices of each character $R_j$ of a string $R$ of $\mathcal{G}_2$ with the same label as $S_i$. If $S_i$ is a cap, $v_{S_i}$ is connected with $v_{R_j}$. Otherwise, $v_{S_i}^t$ is connected with $v_{R_j}^t$ and $v_{S_i}^h$ is connected with $v_{R_j}^h$.

Two vertices $v_{S_i}^t$ and $v_{S_i}^h$ from the same character $S_i$ are called \emph{twin vertices} and two desire edges connecting $v_{S_i}^t$ with $v_{R_j}^t$ and  $v_{S_i}^h$ with $v_{R_j}^h$ are called \emph{twin edges}.

\begin{figure}
    \centering
    \includegraphics[width=0.9\linewidth]{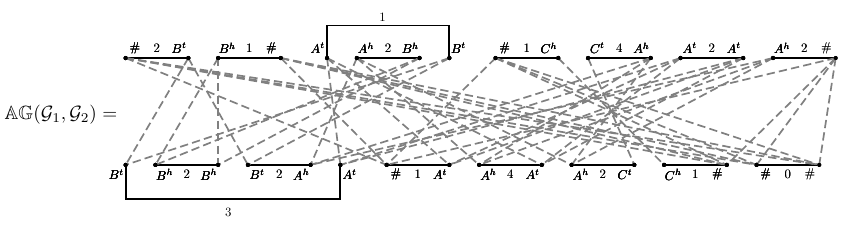}
    \caption{The adjacency graph from the genomes in Figure~\ref{fig:genomes}. The reality edges are shown as continuous lines, whose labels represent the weights, and the desire edges are shown as dashed lines.}
    \label{fig:graph}
\end{figure}

The weight function $w$ is used to include intergenic region information in the graph. For each reality edge $e$ connecting a vertex of $S_i$ with a vertex of $S_{i+1}$, we have $w(e) = \breve{S}_i$. Figure~\ref{fig:graph} shows an adjacency graph created from the genomes in Figure~\ref{fig:genomes}.

\begin{figure}[htb]
    \centering
    \includegraphics[width=0.8\linewidth]{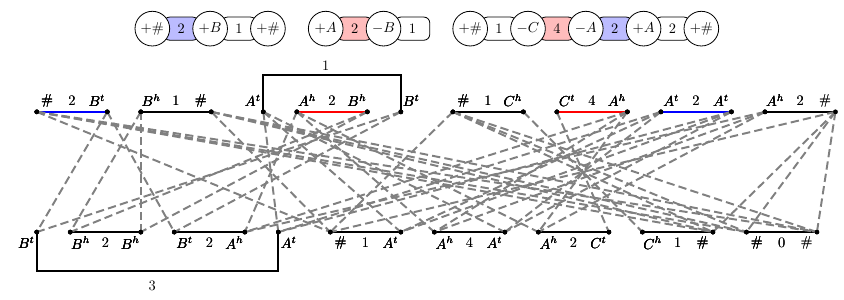}\vspace{1mm}
    \includegraphics[width=0.8\linewidth]{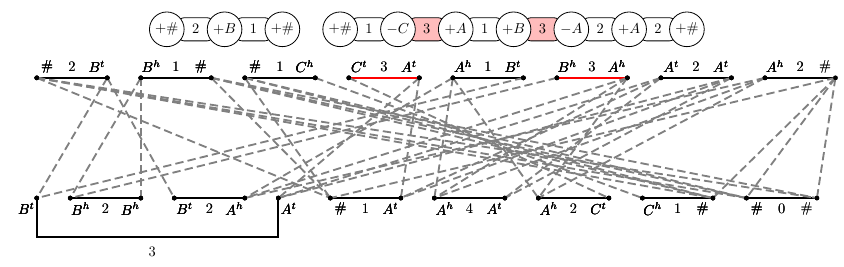}\vspace{1mm}
    \includegraphics[width=0.8\linewidth]{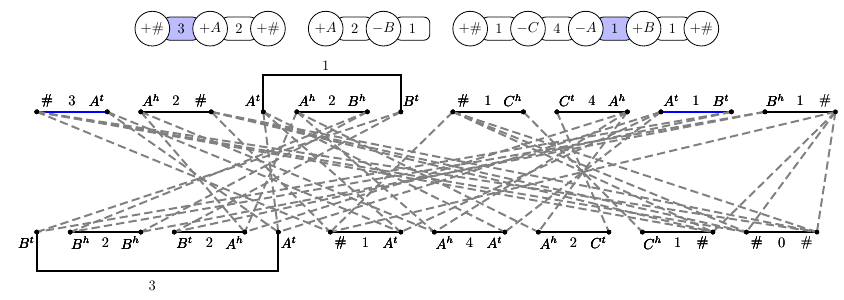}
    \caption{Two applications of the DCJ operation in the genome $\mathcal{G}_1$ of Figure~\ref{fig:genomes} and in the graph of Figure~\ref{fig:graph}. The first operation is $dcj(A^h,B^h,C^t,A^h,1,2)$, the cut and formed intergenic regions and edges are marked in red. The second operation is $dcj(\#,B^t,A^t,A^t,1,0)$, the cut and formed intergenic regions and edges are marked in blue.}
    \label{fig:dcj}
\end{figure}

The \emph{Double Cut and Join (DCJ)} operation can be described using the adjacency graph. Consider four vertices $v_1$, $v_2$, $v_3$ and $v_4$, correspondent to characters of $\mathcal{G}_1$, and two integers $x$ and $y$, such that there is a reality edge $e_1$ connecting $v_1$ with $v_2$ and a reality edge $e_2$ connecting $v_3$ with $v_4$. Besides, we require that $0 \leq x \leq w(e_1)$ and $0 \leq y \leq w(e_2)$. The $dcj(v_1,v_2,v_3,v_4,x,y)$ operation removes the edges $e_1$ and $e_2$ and adds new edges connecting $v_1$ with $v_3$ and connecting $v_2$ with $v_4$. The new edges have weights $x + w(e_2) - y$ and $y + w(e_1) - x$, respectively. This operation can be interpreted as cutting the genome $\mathcal{G}_1$ in two intergenic regions and joining the resulting parts. The points where the genome is cut can be joined in different ways, depending on the order in which the vertices are passed as arguments to $dcj$. The representation of the genome has the signs of the reversed parts flipped, except caps that always have a positive sign. Figure~\ref{fig:dcj} shows two examples of the DCJ operation.

The \emph{DCJ Distance Problem} consists of finding the minimum number of DCJ operations to transform a genome $\mathcal{G}_1$ in another genome $\mathcal{G}_2$.

An \emph{alternating cycle} is a cycle that alternates between reality and desire edges. An alternating cycle is \emph{balanced} if the sum of weights from the reality edges of one of the genomes is equal to the sum of weight from the reality edges of the other genome. An \emph{alternating cycle packing} of the adjacency graph is a set of alternating cycles that do not share edges, contain all vertices, and every desire edge that has a twin edge in a cycle is also in a cycle.

An alternating cycle packing corresponds to an assignment of the genes of $\mathcal{G}_1$ to the genes of $\mathcal{G}_2$ by following the desire edges. With this assignment we can treat the genomes as if they do not have replicated genes. In that case, there is a known $4/3$-approximation algorithm for the DCJ Distance Problem~\cite{2017-fertin-etal}. So we are interested in finding a cycle packing that tries to produce an assignment that leads to the shortest DCJ distance. As the DCJ Distance Problem is NP-hard, we will use the $4/3$-approximation algorithm instead of the exact distance.

\section{Proposed Heuristics}\label{sec:heur}

We are going to adapt the heuristics proposed for a similar cycle packing problem~\cite{2021c-siqueira-etal}, where there are no intergenic regions and each genome has a single linear chromosome. Our heuristics are based on the idea of producing multiple cycle packing and selecting the best one by some criterion. This criterion can be the DCJ distance approximation, but to increase the efficiency of the algorithms we use only the number of balanced cycles in the packing as the selection criteria. The number $r$ of packing to be produced is the same for all heuristics.

The first heuristic, called Simple Random (SR), uses a simple random generation to generate the $r$ packings. At each step, we select a random desire edge that is not the only desire edge incident to its vertices, and remove all other desire edges incident to its vertices. If that edge has a twin we do the same for it.

\begin{figure}[htb]
    \centering
    \includegraphics[width=0.7\linewidth]{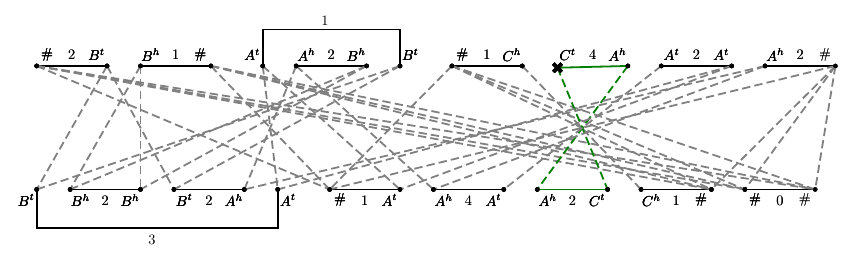}\vspace{1mm}
    \includegraphics[width=0.7\linewidth]{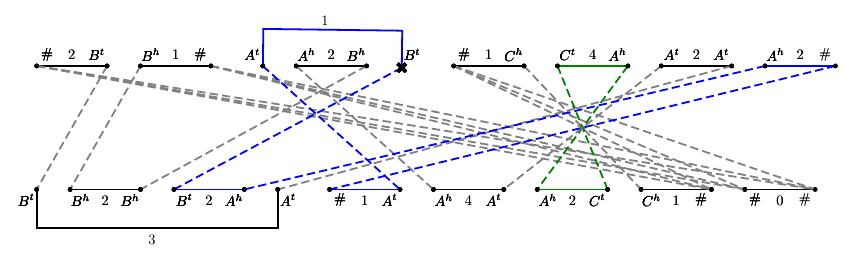}\vspace{1mm}
    \includegraphics[width=0.7\linewidth]{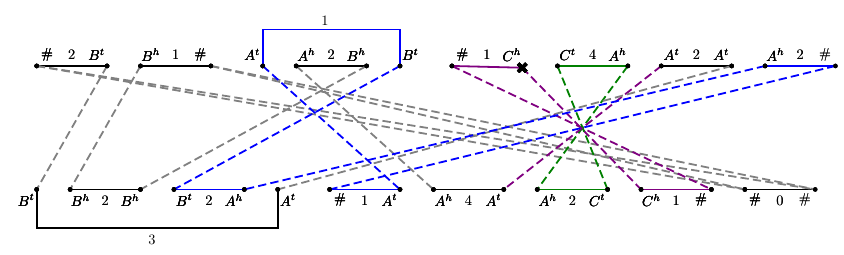}\vspace{1mm}
    \includegraphics[width=0.7\linewidth]{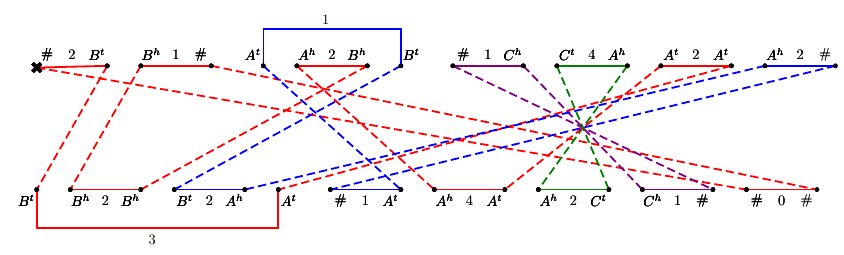}
    \caption{Example of the construction of one packing from the GBFS heuristic applied to the adjacency graph from Figure~\ref{fig:graph}. The cycles obtained on each breadth-first search are marked with an $\times$ indicating the starting vertex of the search. The resulting packing has two balanced cycles.}
    \label{fig:example_gbfs}
\end{figure}

The second heuristic, called Greedy BFS (GBFS), uses a greedy approach by selecting a random vertex and applying a breadth-first search to find the shortest alternating cycle containing that vertex. For every edge selected for the cycle remove all other desire edges incident to its vertices and, if it has a twin, we remove all other desire edges incident to the vertices of this twin. We repeat this process until all vertices are in a cycle. During the breadth-first searches we must keep track of the edges that have to be removed to ensure that the cycle respects the restriction that every edge must have its twin in a cycle as well. Algorithm~\ref{alg:gbfs} shows a pseudo-code of this heuristic, and Figure~\ref{fig:example_gbfs} shows the production of one cycle packing.

\begin{algorithm}[t]
	\caption{Greedy BFS}
	\DontPrintSemicolon%
	\label{alg:gbfs}
	\KwData{A adjacency graph $\mathbb{AG} (\mathcal{G}_1,\mathcal{G}_2)$ and the number of cycle packings $r$}
	\KwResult{A cycle packing for $\mathbb{AG}(\mathcal{G}_1,\mathcal{G}_2)$}
	\BlankLine
	$\mathbf{M} \leftarrow \emptyset$\;
	\While{$|\mathbf{M}| < r$} {%
		$\mathcal{P} \leftarrow \emptyset$\;
		\While{$\mathcal{P}$ does not cover all vertices of $\mathbb{AG}(\mathcal{G}_1,\mathcal{G}_2)$} {%
			$v \leftarrow$ vertex of $\mathbb{AG}(\mathcal{G}_1,\mathcal{G}_2)$ not belonging to any cycle of $\mathbf{M}$\;
			$C \leftarrow$ cycle resulting from a breadth-first search in $\mathbb{AG}(\mathcal{G}_1,\mathcal{G}_2)$, starting with $v$\;
			Remove the necessary vertices from $\mathbb{AG}(\mathcal{G}_1,\mathcal{G}_2)$\;
			$\mathcal{P} \leftarrow \mathcal{P} \cup C$\;
		}
		$\mathbf{M} \leftarrow \mathbf{M} \cup \{\mathcal{P}\}$\;
	}
	\KwRet{Cycle packing belonging to $\mathbf{M}$ with the largest number of balanced cycles}\;
\end{algorithm}

The last heuristic uses a Genetic Algorithm (GA) approach~\cite{1996-mitchell} to produce the set of packings. This approach is inspired by evolution and consists of producing an initial set of individuals, called population, and then applying crossovers and mutations to generate new populations. The individuals are evaluated by a fitness function, which, in our case, is the number of balanced cycles in the packing. The algorithm stops when a fixed number of individuals are generated.

\begin{figure}[htb]
    \centering
    \includegraphics[width=0.75\linewidth]{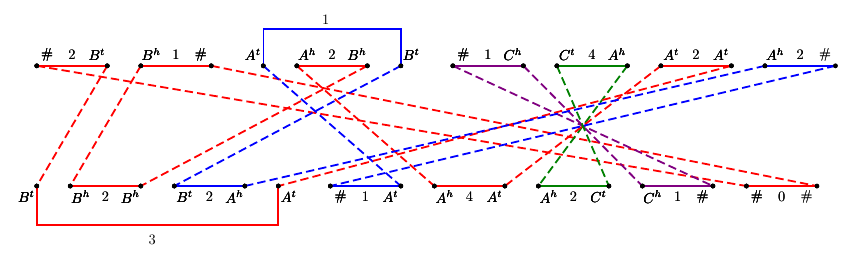}\vspace{1mm}
    \includegraphics[width=0.75\linewidth]{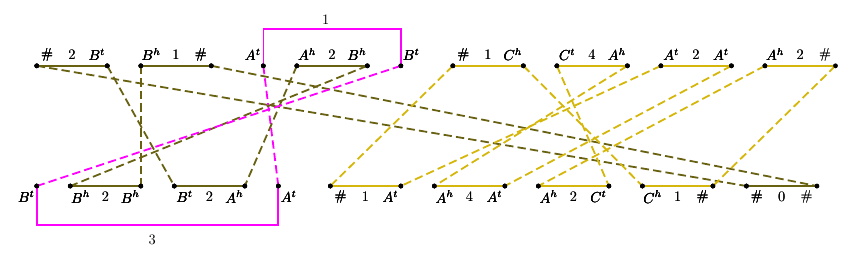}\vspace{1mm}
    \includegraphics[width=0.75\linewidth]{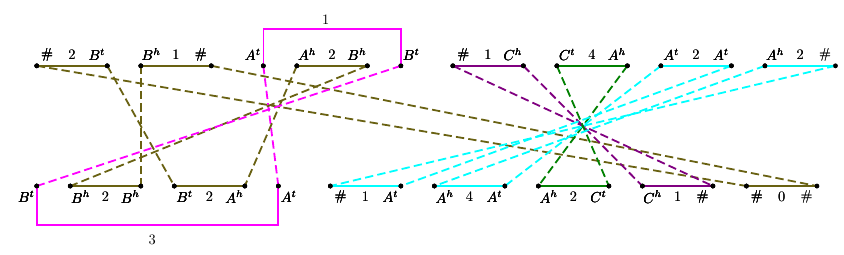}
    \caption{Example of the crossover of two cycle packing of the graph in Figure~\ref{fig:graph}. Two cycles are selected from the first packing, two from the second packing, and one new cycle is inserted by a breadth-first search to complete the packing.}
    \label{fig:cross}
\end{figure}

Besides the total number of individuals $r$, our genetic algorithm is specified by two parameters: the size of the population $p$ ($0 \leq p \leq r$) and the mutation rate $m$ ($0 \leq m \leq 1$). An initial population is generated by the GBFS heuristic. Until a total of $r$ individuals are produced, at each iteration the algorithm produces a new population of $p$ new individuals by applying the following steps:

\begin{itemize}
	\item Selection: At this step, the algorithm selects, with repetition, $2p$ individuals to take part in the crossovers. The selection of each individual is by tournament, in which two individuals are randomly chosen and the one with the highest fitness is selected.
	\item Crossover: The $2p$ selected individuals are paired, and the algorithm applies the crossover operation in each pair to generate a new individual. Given two cycle packings $\mathcal{P}$ and $\mathcal{P}'$ of an adjacency graph, a \emph{crossover} of $\mathcal{P}$ and $\mathcal{P}'$ is a new cycle packing created by the following procedure. Let $L$ and $L'$ be two randomly ordered lists of the cycles from $\mathcal{P}$ and $\mathcal{P}'$, respectively. Starting with an empty set $\mathcal{P}''$ and with the original adjacency graph, while $\mathcal{P}''$ is not a cycle packing and there are cycles in $L$ or $L'$, remove a cycle from $L$ or $L'$ (with a 50\% probability to remove from each list, or 100\% probability to remove from one list if the other is empty), and add it to $\mathcal{P}''$ if all its edges are still available. As in the GBFS heuristic, remove the necessary edges from the graph to ensure the restriction about the twin edges. If both lists $L$ and $L'$ are empty and $\mathcal{P}''$ is not yet a cycle packing, use breadth-first searches, as in the GBFS heuristic, to complete the packing. Figure~\ref{fig:cross} shows an example of crossover.
	\item Mutation: After the new individuals are generated, a mutation is applied to each one in order to increase the diversity of the population. In the mutation of a cycle pack $\mathcal{P}''$, we remove each cycle of $\mathcal{P}''$ with probability $m$ and create a new packing from the remaining cycles of $\mathcal{P}''$ using breadth-first searches.
	\item Elitism: As we are going to replace the old population with the new one, we cannot guarantee the quality of the new population. So, to ensure that at least the best individual is kept, we replace the individual with the lowest fitness from the new population with the individual with the highest fitness from the old population, if the old individual has a higher fitness than the new one.
\end{itemize}

Algorithm~\ref{alg:ga} shows a pseudo-code of the GA heuristic.

\begin{algorithm}[t]
	\caption{Genetic Algorithm}
	\DontPrintSemicolon%
	\label{alg:ga}
	\KwData{An adjacency graph $\mathbb{AG} (\mathcal{G}_1,\mathcal{G}_2)$, the number of cycle packings $r$, size of the population $p$, and mutation rate $m$}
	\KwResult{An adjacency packing for $\mathbb{AG}(\mathcal{G}_1,\mathcal{G}_2)$}
	\BlankLine
	$\mathbf{P}_0 \leftarrow p$ cycle packings generated by the GBFS heuristic\;
	$\mathbf{M} \leftarrow \mathbf{P}_0$\;
	$i \leftarrow 0$\;
	\While{$|\mathbf{M}| < r$} {%
		$\mathbf{S} \leftarrow$ Sequence of $p$ selected pairs of individuals from $\mathbf{P}_i$\;
		$\mathbf{P}_{i+1}' \leftarrow$ Set of $p$ new individuals created by crossover of individuals from $\mathbf{S}$\;
		$\mathbf{P}_{i+1}'' \leftarrow$ Set of $p$ individuals obtained by mutation of individuals from $\mathbf{P}_{i+1}'$\;
		\eIf {the best individual of $P_{i}$ is better than the worse individual of $P_{i+1}''$}{%
			$\mathbf{P}_{i+1} \leftarrow$ Set $\mathbf{P}_{i+1}''$ with its worse individual replaced with the best individual of $P_{i}$\;
		}{%
			$\mathbf{P}_{i+1} \leftarrow \mathbf{P}_{i+1}''$\;
		}
		$\mathbf{M} \leftarrow \mathbf{M} \cup \{\mathbf{P}_{i+1}\}$\;
		$i \leftarrow i+1$\;
	}
	\KwRet{Cycle packing belonging to $\mathbf{M}$ with the largest number of balanced cycles}\;
\end{algorithm}

\section{Experiments}\label{sec:exp}

The instances used in our experiments and the implementation of the heuristics (in C++) are available at a public repository\footnote{https://github.com/compbiogroup/Heuristics-based-on-Adjacency-Graph-Packing-for-DCJ-Distance-Considering-Intergenic-Regions}. The following tests were conducted on a computer equipped with an ``Intel(R) Xeon(R) CPU E5-2470 v2'' with 40 cores at 2.40GHz, 32GB of RAM, and 9GB of swap space.

We created a database of simulated genomes to test our heuristics. We produced $25000$ pair of genomes separated into groups of $1000$. Each group is defined by the number $\ell$ of labels used and the number $o$ of DCJ operations applied. We created each pair by the following process (each random selection is uniformly distributed):
\begin{itemize}
	\item We randomly produced $200$ characters picking the label from a set with $\ell$ symbols and choosing the signs randomly.
	\item We randomly chose $201$ integers in the interval $[0,100]$ to represent the intergenic regions.
	\item We produced a genome $\mathcal{G}$ containing a single linear chromosome with the chosen characters and intergenic regions.
	\item We created the first genome of the pair by applying $o/2$ DCJ operations in $\mathcal{G}$.
	\item We created the second genome of the pair by applying $o/2$ DCJ operations in $\mathcal{G}$.
\end{itemize}

For our tests all heuristics produced a set of $r=1000$ alternating cycle packings. For the GA heuristic, we chose the parameters $m=0.2$ and $p=100$.

In Table~\ref{tab:dcj} and Table~\ref{tab:time} we can see the results of our experiments. For each heuristic and each group of instances, we show, in Table~\ref{tab:dcj}, the average and standard deviation of the distance computed from the best cycle packing found and, in Table~\ref{tab:time}, the average and standard deviation of the execution time (n seconds) of the algorithms. {\REVIEW In both tables the standard deviation is shown as a percentage of the mean.}

We can see that GA found the shortest distance on average, followed by GBFS and SR. That indicates that using a more sophisticated strategy to produce the packings results in a better result. The largest difference between the average distances found by GA and by GBFS is $54.44$ for $o=40$ and $\ell=40$, while the shortest such difference is $6.29$ for $o=10$ and $\ell=50$. Considering the standard deviation, the distinction between the distances found by GA and by GBFS is clearer with larger values of $o$.

Additionally, the running time of GA was shorter than the running time of the other heuristics. This is a consequence of the fact that the GA heuristic does not produce all cycle packings from scratch, but produces most packings from crossovers and mutations.

\newcolumntype{g}{>{\columncolor{Gray}}r}
\newcolumntype{G}{>{\columncolor{Gray}}c}
\begin{table*}[htb]
	\centering
	\caption{Average and standard deviation for the DCJ distances resulting from the heuristics.}
	\label{tab:dcj}
	{
		\def\arraystretch{1.1}\tabcolsep=4pt
		\begin{tabular}{rrgGgrcrgGg}
			\hline
			$o$ & $\ell$ & \multicolumn{3}{G}{\bf GA} & \multicolumn{3}{c}{\bf GBFS} & \multicolumn{3}{G}{\bf SR}                                \\
			\hhline{-|-|-|-|-|-|-|-|-|-|-|}
			$10$ & $10$ & $145.03$ & $\pm$ & 15.21\% & $172.02$ & $\pm$ & 10.86\% & $204.44$ & $\pm$ & 11.14\%\\
			$10$ & $20$ & $59.03 $ & $\pm$ & 51.18\% & $108.45$ & $\pm$ & 35.98\% & $148.47$ & $\pm$ & 33.12\%\\
			$10$ & $30$ & $25.60 $ & $\pm$ & 66.17\% & $51.23 $ & $\pm$ & 64.45\% & $87.31 $ & $\pm$ & 63.51\%\\
			$10$ & $40$ & $16.15 $ & $\pm$ & 63.07\% & $27.74 $ & $\pm$ & 75.39\% & $54.80 $ & $\pm$ & 86.22\%\\
			$10$ & $50$ & $12.01 $ & $\pm$ & 44.94\% & $18.30 $ & $\pm$ & 73.85\% & $35.55 $ & $\pm$ & 99.68\%\\
			\hhline{-|-|-|-|-|-|-|-|-|-|-|}
			$20$ & $10$ & $163.85$ & $\pm$ &  9.80\% & $186.75$ & $\pm$ & 6.25\% & $214.11$ & $\pm$ & 5.70\%\\
			$20$ & $20$ & $103.75$ & $\pm$ & 28.18\% & $152.03$ & $\pm$ & 14.79\% & $182.64$ & $\pm$ & 13.58\%\\
			$20$ & $30$ & $59.35 $ & $\pm$ & 38.88\% & $109.22$ & $\pm$ & 30.38\% & $146.76$ & $\pm$ & 26.82\%\\
			$20$ & $40$ & $37.29 $ & $\pm$ & 43.65\% & $75.06 $ & $\pm$ & 43.67\% & $109.62$ & $\pm$ & 41.23\%\\
			$20$ & $50$ & $28.58 $ & $\pm$ & 37.72\% & $52.02 $ & $\pm$ & 49.17\% & $81.83 $ & $\pm$ & 52.58\%\\
			\hhline{-|-|-|-|-|-|-|-|-|-|-|}
			$30$ & $10$ & $175.16$ & $\pm$ & 7.61\% & $194.90$ & $\pm$ & 5.45\% & $217.66$ & $\pm$ & 4.02\%\\
			$30$ & $20$ & $135.22$ & $\pm$ & 17.37\% & $172.33$ & $\pm$ & 8.77\% & $196.27$ & $\pm$ & 7.93\%\\
			$30$ & $30$ & $96.39 $ & $\pm$ & 27.27\% & $147.38$ & $\pm$ & 14.37\% & $172.66$ & $\pm$ & 13.77\%\\
			$30$ & $40$ & $68.36 $ & $\pm$ & 34.96\% & $120.39$ & $\pm$ & 22.65\% & $149.94$ & $\pm$ & 19.43\%\\
			$30$ & $50$ & $51.12 $ & $\pm$ & 32.35\% & $93.15 $ & $\pm$ & 30.95\% & $123.60$ & $\pm$ & 29.17\%\\
			\hhline{-|-|-|-|-|-|-|-|-|-|-|}
			$40$ & $10$ & $185.00$ & $\pm$ & 5.75 \% & $201.67$ & $\pm$ & 4.20\% & $219.28$ & $\pm$ & 3.38\%\\
			$40$ & $20$ & $156.60$ & $\pm$ & 10.76\% & $184.72$ & $\pm$ & 6.37\% & $203.64$ & $\pm$ & 5.53\%\\
			$40$ & $30$ & $119.66$ & $\pm$ & 18.99\% & $165.50$ & $\pm$ & 9.60\% & $186.53$ & $\pm$ & 8.91\%\\
			$40$ & $40$ & $92.53 $ & $\pm$ & 24.36\% & $146.97$ & $\pm$ & 13.69\% & $169.63$ & $\pm$ & 12.04\%\\
			$40$ & $50$ & $75.16 $ & $\pm$ & 26.47\% & $127.49$ & $\pm$ & 17.35\% & $151.83$ & $\pm$ & 16.10\%\\
			\hhline{-|-|-|-|-|-|-|-|-|-|-|}
			$50$ & $10$ & $191.97$ & $\pm$ &  4.72\% & $205.41$ & $\pm$ & 3.74\% & $221.23$ & $\pm$ & 2.86\%\\
			$50$ & $20$ & $168.78$ & $\pm$ &  7.91\% & $192.35$ & $\pm$ & 5.03\% & $208.59$ & $\pm$ & 4.08\%\\
			$50$ & $30$ & $144.03$ & $\pm$ & 12.30\% & $178.56$ & $\pm$ & 7.17\% & $195.46$ & $\pm$ & 6.28\%\\
			$50$ & $40$ & $117.53$ & $\pm$ & 18.38\% & $163.73$ & $\pm$ & 9.80\% & $182.52$ & $\pm$ & 8.61\%\\
			$50$ & $50$ & $98.91 $ & $\pm$ & 21.72\% & $147.36$ & $\pm$ & 12.51\% & $168.67$ & $\pm$ & 11.39\%\\
			\hline
		\end{tabular}
	}
\end{table*}

\begin{table*}[htb]
	\centering
	\caption{Average and standard deviation for the running times of the heuristics (in seconds).}
	\label{tab:time}
	{
		\def\arraystretch{1.1}\tabcolsep=4pt
		\begin{tabular}{rrgGgrcrgGg}
			\hline
			$o$ & $\ell$ & \multicolumn{3}{G}{\bf GA} & \multicolumn{3}{c}{\bf GBFS} & \multicolumn{3}{G}{\bf SR}                                \\
			\hhline{-|-|-|-|-|-|-|-|-|-|-|}
			10 & 10 & 36.78 & $\pm$ & 4.11\% & 213.56 & $\pm$ & 26.05\% & 40.78 & $\pm$ & 14.53\% \\
			10 & 20 & 25.73 & $\pm$ & 7.41\% & 136.13 & $\pm$ & 41.69\% & 40.61 & $\pm$ & 13.82\% \\
			10 & 30 & 16.10 & $\pm$ & 6.94\% & 67.38  & $\pm$ & 37.00\% & 38.25 & $\pm$ & 13.56\% \\
			10 & 40 & 11.88 & $\pm$ & 5.58\% & 37.31  & $\pm$ & 28.46\% & 37.52 & $\pm$ & 12.81\% \\
			10 & 50 & 10.31 & $\pm$ & 4.05\% & 22.75  & $\pm$ & 19.60\% & 34.95 & $\pm$ & 11.60\% \\
			\hhline{-|-|-|-|-|-|-|-|-|-|-|}
			20 & 10 & 37.91 & $\pm$ & 3.46\% & 225.08 & $\pm$ & 23.60\% & 39.70 & $\pm$ & 14.25\% \\
			20 & 20 & 32.76 & $\pm$ & 6.35\% & 172.17 & $\pm$ & 35.19\% & 39.64 & $\pm$ & 14.05\% \\
			20 & 30 & 26.04 & $\pm$ & 8.09\% & 119.97 & $\pm$ & 44.08\% & 39.62 & $\pm$ & 13.48\% \\
			20 & 40 & 19.13 & $\pm$ & 8.04\% & 82.09  & $\pm$ & 43.96\% & 38.43 & $\pm$ & 12.97\% \\
			20 & 50 & 15.39 & $\pm$ & 7.88\% & 56.28  & $\pm$ & 38.76\% & 36.28 & $\pm$ & 11.96\% \\
			\hhline{-|-|-|-|-|-|-|-|-|-|-|}
			30 & 10 & 38.07 & $\pm$ & 3.45\% & 232.65 & $\pm$ & 24.05\% & 41.24 & $\pm$ & 14.62\% \\
			30 & 20 & 37.30 & $\pm$ & 5.73\% & 195.65 & $\pm$ & 29.77\% & 40.57 & $\pm$ & 14.26\% \\
			30 & 30 & 33.24 & $\pm$ & 7.82\% & 164.24 & $\pm$ & 40.45\% & 40.24 & $\pm$ & 13.33\% \\
			30 & 40 & 29.02 & $\pm$ & 9.58\% & 132.96 & $\pm$ & 47.19\% & 39.22 & $\pm$ & 13.39\% \\
			30 & 50 & 24.13 & $\pm$ & 9.82\% & 102.31 & $\pm$ & 49.57\% & 37.05 & $\pm$ & 12.37\% \\
			\hhline{-|-|-|-|-|-|-|-|-|-|-|}
			40 & 10 & 39.03 & $\pm$ & 3.38\% & 240.62 & $\pm$ & 21.66\% & 41.05 & $\pm$ & 14.24\% \\
			40 & 20 & 39.48 & $\pm$ & 4.90\% & 202.19 & $\pm$ & 29.20\% & 40.69 & $\pm$ & 14.11\% \\
			40 & 30 & 38.01 & $\pm$ & 6.84\% & 180.91 & $\pm$ & 36.32\% & 40.47 & $\pm$ & 14.18\% \\
			40 & 40 & 35.15 & $\pm$ & 8.67\% & 160.33 & $\pm$ & 45.16\% & 40.09 & $\pm$ & 13.15\% \\
			40 & 50 & 31.70 & $\pm$ & 9.77\% & 139.39 & $\pm$ & 47.48\% & 38.39 & $\pm$ & 12.49\% \\
			\hhline{-|-|-|-|-|-|-|-|-|-|-|}
			50 & 10 & 39.62 & $\pm$ & 3.54\% & 243.66 & $\pm$ & 20.86\% & 41.44 & $\pm$ & 14.56\% \\
			50 & 20 & 40.99 & $\pm$ & 4.98\% & 211.95 & $\pm$ & 28.17\% & 39.90 & $\pm$ & 13.85\% \\
			50 & 30 & 41.46 & $\pm$ & 6.53\% & 194.58 & $\pm$ & 36.40\% & 40.08 & $\pm$ & 13.76\% \\
			50 & 40 & 39.86 & $\pm$ & 8.21\% & 180.37 & $\pm$ & 43.45\% & 40.55 & $\pm$ & 13.47\% \\
			50 & 50 & 38.77 & $\pm$ & 9.86\% & 161.15 & $\pm$ & 46.69\% & 39.22 & $\pm$ & 13.08\% \\
			\hline
		\end{tabular}
	}
\end{table*}

\section{Conclusion}\label{sec:conc}

We described and tested three heuristics for Adjacency Graph Packing that can be used to estimate DCJ distances. One of the heuristics is based on a simple random strategy (SR), another uses a greedy strategy based on BFS (GBFS), and the last one is based on Genetic Algorithms (GA). These heuristics were tested on a database of simulated genomes and GA found the shorter distances on average with the shortest running time.

This work can be further extended to consider genomes with distinct sets of genes, as was done for the original heuristics developed for unichromosomal genomes without intergenic regions~\cite{2024-siqueira-etal}. There are known algorithms for rearrangement distances with intergenic regions and distinct sets of genes, but without gene replicas~\cite{2021b-alexandrino-etal,2021c-alexandrino-etal}. It is also relevant to explore other approaches to the problem, including exact and approximation algorithms. The use of String Partition problems with intergenic regions can be one approach to develop approximations~\cite{2021b-siqueira-etal, 2022-siqueira-etal}.

{\REVIEW
Another path for future works is the application of the proposed algorithms to compare real genomes. These tests can include the integration of the algorithms in bioinformatic tools used in the construction of phylogenetic trees or detection of orthologous genes.
}

\section*{Acknowledgement}
This work was supported by the S\~ao Paulo Research Foundation, FAPESP (grant %
2021/13824-8
).

\bibliographystyle{sbc}
\bibliography{main}

\end{document}